\documentclass[aps, preprint, superscriptaddress]{revtex4-1}
\usepackage{graphicx, amsmath, amssymb, physics}
\usepackage{xcolor}

\begin{document}

\title{Random Singlet Phase of Cold Atoms Coupled to a  Photonic Crystal Waveguide}

\date{\today}

\author{David Z. Li}
\affiliation{ICFO-Institut de Ciencies Fotoniques, The Barcelona Institute of Science and Technology, 08860 Castelldefels (Barcelona), Spain}
\author{Marco T. Manzoni}
\affiliation{ICFO-Institut de Ciencies Fotoniques, The Barcelona Institute of Science and Technology, 08860 Castelldefels (Barcelona), Spain}
\author{Darrick E. Chang}
\affiliation{ICFO-Institut de Ciencies Fotoniques, The Barcelona Institute of Science and Technology, 08860 Castelldefels (Barcelona), Spain}
\affiliation{ICREA-Instituci\'{o} Catalana de Recerca i Estudis Avan\c{c}ats, 08015 Barcleona, Spain}

\begin{abstract}
Systems consisting of cold atoms trapped near photonic crystal waveguides have recently emerged as an exciting platform for quantum atom-light interfaces. Such a system enables realization of tunable long-range interactions between internal states of atoms (spins), mediated by guided photons. Currently, experimental platforms are still limited by low filling fractions, where the atom number is much smaller than the number of sites at which atoms can potentially be trapped. Here, we show that this regime in fact enables interesting many-body quantum phenomena, which are typically associated with short-range disordered systems. As an example, we show how the system can realize the so-called ``random singlet phase", in which all atoms pair into entangled singlets, but the pairing occurs over a distribution of ranges as opposed to nearest neighbors. We use a renormalization group method to obtain the distribution of spin entanglement in the random singlet phase, and show how this state can be approximately reached via adiabatic evolution from the ground state of a non-interacting Hamiltonian. We also discuss how experimentally this random singlet phase can be observed.
We anticipate that this work will accelerate the route toward the exploration of strongly correlated matter in atom-nanophotonics interfaces, by avoiding the requirement of perfectly filled lattices.

\end{abstract}

\pacs{xxxx}


\maketitle

\textit{Introduction} 
In recent years, there has been considerable effort in interfacing atoms and other quantum emitters with nanophotonic structures \cite{Lodahl2015, Chang2018}, including nanofibers \cite{Vetsch2010, Goban2012, Mitsch2014, Sorensen2016, Corzo2016, Corzo2019, Kato2019} and photonic crystal waveguides (PCWs) \cite{Lund2008, Javadi2015, Goban2014, Hood2016, Kim2019}. The predominant aim of such efforts initially was to utilize the potentially strong light-matter interactions in such systems, arising from the nanoscale confinement of optical fields, for applications within quantum information processing  \cite{Kimble2008, Oshea2013, Tiecke2014, Shomroni2014, Scheucher2016}. More recently, however, it has been realized that these atom-nanophotonics interfaces also open up new paradigms to explore quantum many-body physics \cite{Douglas2015, Gonzalez2015, Hung2016, Manzoni2017, Chang2018, Prasad2019, Mahmoodian2019}.

\par
In particular, when an atomic transition frequency lies in a bandgap of a PCW, a photon emitted from an atom becomes an evanescent wave and forms  a bound state around the atom. Multiple atoms coupled to a PCW can exchange excitations via these localized photons, giving rise to effective spin interactions whose range is determined by the decay length of the evanescent wave, which in turn can be tuned via the detuning between the atomic transition frequency and the band edge of the photonic crystal \cite{Kurizki1990, John1996, Bay1997, Shahmoon2013, Douglas2015, Gonzalez2015}. Theoretically, there has been interest in using atom-PCW interfaces to investigate long-range spin models \cite{Hung2016, Liu2019}, strong spin-motion coupling \cite{Manzoni2017}, or long-range interactions between photons \cite{Douglas2016, Shahmoon2016}. 

\par
These proposals typically require perfect filling of the lattice sites where atoms can potentially be trapped, which is a challenge in current experiments \cite{Goban2014, Hood2016}. Here, we show that the combination of long-range interactions and low filling enables the realization of novel many-body physics, allowing the system to mimick a spin chain with short-range, random interaction strength \cite{Dasgupta, Fisher1994}. In particular, under certain conditions, the ground state of the system becomes a ``random singlet phase," where all atoms entangle into singlet pairs, but the pairing occurs over a distribution of ranges instead of between nearest neighbors. We analyze the main properties of this phase, and discuss how it can be prepared and observed in a realistic PCW system.

\begin{figure}[h]
\centering
\includegraphics[scale=0.6]{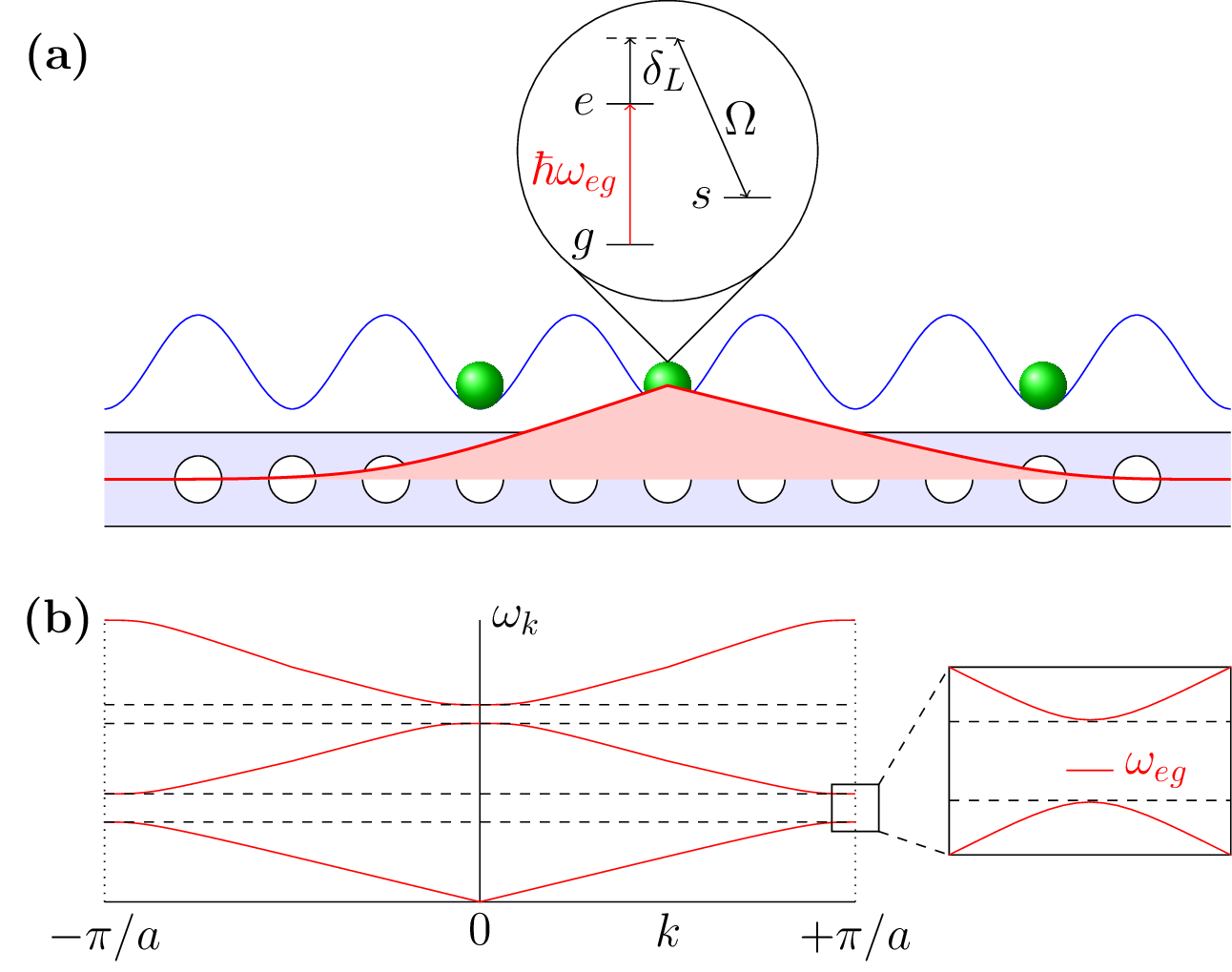}
\caption{\textbf{(a)} Schematic illustration of setup, consisting of a sparse and random filling of cold atoms (green dots) tightly trapped in a lattice potential (blue periodic curve) near a 1D PCW. The atoms have a $\Lambda$-level scheme, with the ground ($g$) to excited ($e$) state transition frequency being $\omega_{eg}$, and the excited and a metastable state ($s$) coupled by a Raman laser with Rabi frequency $\Omega$ and detuning $\delta_L$. If $\omega_{eg}$ lies in the bandgap of the PhC, a photon bound state can be formed around the atom, illustrated as the red decay envelope. \textbf{(b)} Typical band structure of a 1D PhC with bandgaps, with guided mode frequency $\omega_k$ as a function of Bloch wavevector $k$. The zoom-in rectangle shows the atomic transition frequency $\omega_{eg}$ situated in a bandgap and close to a bandedge.}
\label{fig:PhC}
\end{figure}

A photonic crystal is a periodic dielectric structure that controls the propagation of light. 
Due to the periodicity, the dispersion relation $\omega_k$ versus Bloch wavevector $k$ of guided modes is describable by bands (Fig. \ref{fig:PhC}\textbf{(b)}). We assume that the atomic optical transition (involving ground state $\ket{g}$ and excited state $\ket{e}$) is situated within a bandgap, a frequency window in which no propagating modes exist.
This prevents an excited atom from decaying into $\ket{g}$ by emitting a guided photon; however, the state $\ket{e}$ can become dressed by a photon bound state localized a distance $L$ around the atom (Fig. \ref{fig:PhC}\textbf{(a)}). 
Given a second atom in its ground state within a distance $\sim L$ of the first, the pair can exchange their excitations via the bound photon, resulting in an effective spin interaction.
In practice, to avoid the typically fast spontaneous emission rate of $\ket{e}$ into free space, and to also allow the interaction strength to be time-dependent, it is convenient to introduce an additional metastable state $\ket{s}$, which is coupled to $\ket{e}$ via an external laser field with Rabi frequency $\Omega(t)$ (see Fig. \ref{fig:PhC}(a)). Under certain conditions \cite{Douglas2015}, the state $\ket{e}$ and its photon bound state are only virtually excited, allowing the dynamics to be projected into the $\left\{\ket{g}, \ket{s}\right\}$ manifold with effective Hamiltonian 
\begin{equation}\label{eq:HI}
\hat{H}_{int}^N=\sum_{i<j}\hat{H}_{i j}= (1/2)\sum_{i<j} J_{ij}(t)(\hat{\sigma}_{x}^i\hat{\sigma}_{x}^j + \hat{\sigma}_{y}^i\hat{\sigma}_{y}^j) \, ,
\end{equation}
where $\hat{H}_{i j}= (J_{ij}(t)/2) (\hat{\sigma}_{x}^i\hat{\sigma}_{x}^j + \hat{\sigma}_{y}^i\hat{\sigma}_{y}^j)$ denotes the spin-flip pair interaction between atoms $i$ and $j$  ($ i, j = 1, ..., N$ and $N$ is the total number of atoms), with $\left\{\ket{g}, \ket{s}\right\}$ being treated as pseudo-spins $\left\{\ket{\uparrow}, \ket{\downarrow}\right\}$ and $\left\{\hat{\sigma}_{x}, \hat{\sigma}_{y}\right\}$ the regular Pauli matrices. $J_{ij}(t)=J_0(t) \, \exp(-|x_i-x_j|/L)$ where $J_0(t) \propto \Omega(t)$ is a tunable interaction strength proportional to the external field, $L$ is the range of interaction, and $x_i$, $x_j$ are the positions of atoms $i$ and $j$. When the atoms are trapped in discrete positions (corresponding to integer multiples of the  lattice constant), but fill only a small fraction of all possible sites, the distances $|x_i-x_j|$ and the interaction strengths $J_{ij}$ become random (over a set of possible discrete values). The dependence of $J_0$ and $L$ on system parameters (such as laser detunings, band edge curvature, etc.) are detailed elsewhere \cite{Douglas2015}, but not of paramount importance here.

\begin{figure}[h]
\centering
\includegraphics[scale=0.6]{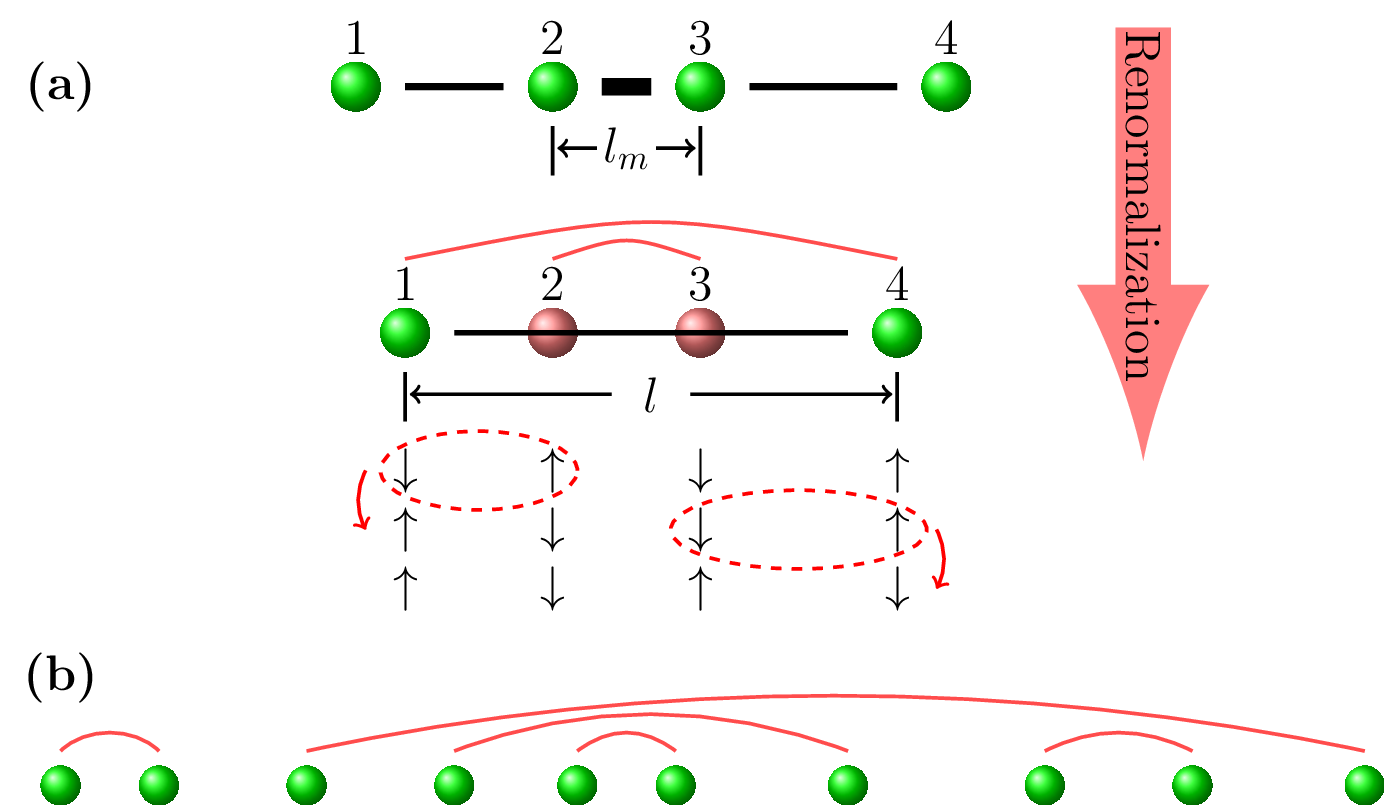}
\caption{\textbf{(a)}: Illustration of renormalization process and interactions mediated by singlet pairs, for the case of four atoms. 
In the top line, atoms 2 and 3 experience the strongest interaction (thick black line) due to their proximity,
so they form a singlet pair (red line) and can be ``frozen out" of the 1D chain (transparent red balls). Below, we indicate half of the singlet state, with atom 2 initially in state $\ket{\uparrow}$ and atom 3 in state $\ket{\downarrow}$. (First line) If atoms 1 and 4 are in states $\ket{\downarrow}$ and $\ket{\uparrow}$, respectively, atoms 1 and 2 can virtually exchange their spins (dashed red circle) at a high energy cost. Atoms 2 and 3 can return to the singlet state (third line) if 3 and 4 virtually exchange their spins (dashed red circle) as well (second line). The entire process overall results in an effective interaction between atom 1 and 4 mediated by the singlet pair of atom 2 and 3. The effective distance $l$ between atom 1 and 4 is therefore ``renormalized" and shrunk by an amount of $d_{\text{eff}}$ (expression given in the text). This procedure generates singlet pairs sitting inside longer ones, which we call ``nesting". 
\textbf{(b)}:  A representative ground state of the \textit{random singlet phase} for ten atoms.}
\label{fig:med}
\end{figure}

We first discuss the properties of the ground state of Eq. (\ref{eq:HI}), in the case that $J_{ij}$ is time-independent, before discussing its preparation by adiabatic evolution. The salient properties of the ground state can be obtained using the renormalization group procedure  introduced in refs. \cite{Dasgupta, Fisher1994}. 
In particular, given a pair of atoms (say $i$ and $i+1$) separated by the shortest distance (denoted as $l_m<l$, where $l$ is any other coupling distance in the system, see the first row in Fig. \ref{fig:med}\textbf{(a)}), and thus experiencing the strongest interaction, we first diagonalize the system around $\hat{H}_{i,i+1}$ and treat the rest of Eq. (\ref{eq:HI}) as a perturbation. For positive $J_0$, the ground state of $\hat{H}_{i,i+1}$ is a singlet: $\ket{S}=\left(1/\sqrt{2}\right) \left(\ket{\uparrow}_{i}\ket{\downarrow}_{i+1} - \ket{\downarrow}_{i}\ket{\uparrow}_{i+1} \right)$.
A spin-flip interaction of one of these atoms (say $i$) with another atom $j\neq i, i+1$ would bring the pair out of the singlet state, at a high energy cost. However, through a second order process, atom $i+1$ can interact with atom $j'\neq j, i, i+1$, which brings the pair back to the singlet and results in an effective spin flip interaction between atoms $j$ and $j'$.
 Remarkably, the new total effective Hamiltonian $H^{N-2}_{int}$ for the remaining $N-2$ atoms takes exactly the same form as Eq. (\ref{eq:HI}), but where the distance between atoms on \textit{opposite} sides of the already paired atoms ($i$ and $i+1$)
 is shortened or \textit{renormalized} (see Fig. \ref{fig:med}):
$\tilde{J}_{j j'} \rightarrow J_0 \, \exp[-(|x_j-x_{j'}|- d_{\text{eff}})/L]$,
where $d_{\text{eff}}/L=2\, l_m/L+\ln\left(1-2e^{-l_m/L}+2e^{-2 \, l_m/L} \right)$, and the new ``effective distance" between atoms $j$ and $j'$ becomes $l \equiv |x_j-x_{j'}|-d_{\text{eff}}$ (see the second row in Fig. \ref{fig:med}\textbf{(a)}). The interactions between atoms on the \textit{same} side of the pair remain roughly unchanged (see Appendix). Thus, the interaction between two atoms on opposite sides of the singlet pair becomes stronger due to the mediating effect of the pair.
One can then repeat this argument, progressively eliminating the next strongest interacting pair with correspondingly larger $l_m$.
 The final result is a many-body ground state composed of only singlet pairings, but the pairing does not necessarily occur between nearest neighbors (Fig. \ref{fig:med}\textbf{(b)}): the so-called \textit{random singlet phase} \cite{Fisher1994}.

\begin{figure}[h]
\centering
\includegraphics[clip=true, trim=2cm 0cm 2cm 0cm, scale=0.4]{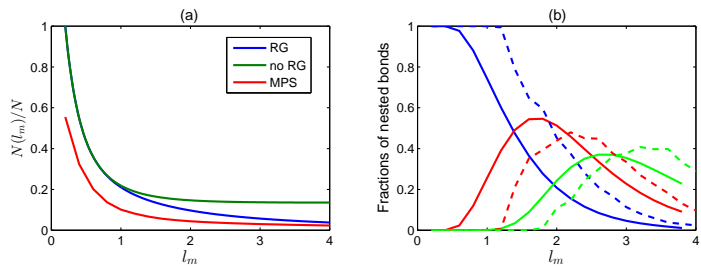}
\caption{\textbf{(a)} Fraction of atoms $N(l_m)/N$ left unpaired, after pairs of atoms with an effective interaction distance up to $l_m$ have been renormalized into singlets. The blue and red curves denote the results predicted by the RG flow equations and by numerical MPS simulations, respectively. For comparison, the curve in green denotes the unpaired atoms without RG, \textit{i.e.} without allowing nested pairs to occur. \textbf{(b)} Among the paired atoms, we plot the fraction of nested bonds at $l_m$ at 30\% filling. The solid and dashed lines are for RG and MPS simulated results respectively. The blue, red and green colors represent nesting of order 
 $n_{l_m} = 0, 1, 2$ respectively.}
\label{fig:MPS}
\end{figure}

To quantify the salient properties of the random singlet phase, one can consider the probability density $P(l,l_m)$, where $ P(l,l_m) \, \text{d}l$ characterizes the probability of finding nearest, unpaired atoms with an effective interaction strength between 
$J_0 \exp(-l/L)$ and $J_0 \exp(-(l+\text{d}l)/L)$, after all pairs interacting with an effective distance of $l_m$ or less have been frozen into singlets. 
Instead of working with $P(l,l_m)$ directly, it is more convenient to perform a change of variables to $Q(\lambda, l_m)=l_m P(l,l_m)$ with $\lambda=l/l_m-1$. Then, it can be shown (see Appendix) that the elimination process
 results in the following evolution or RG flow equation for $Q(\lambda, l_m)$:
\begin{multline}\label{eq:flow}
-Q(\lambda,l_m)+l_m\frac{\partial Q}{\partial l_m} - (\lambda + 1)\frac{\partial Q}{\partial \lambda} =\\
Q(0,l_m)\int_0^{\lambda+g(l_m)} \text{d}\lambda_1 Q(\lambda_1,l_m)Q(\lambda+g(l_m)-\lambda_1,l_m) \, ,
\end{multline}
where $l_m g(l_m)=\ln\left[ 1 - 2e^{-l_m}\left( 1 - e^{-l_m}\right)\right]$ (both $l$ and $l_m$ have been rescaled by $L$). We solve  Eq. (\ref{eq:flow}) numerically, and in Fig. \ref{fig:MPS}\textbf{(a)} show the result for the fraction of atoms remaining unpaired as a function of $l_m$ obtained from the solution of $Q(\lambda, l_m)$, taking example parameters of 30\% filling and an interaction range of $L=5a$.
As expected, as the renormalization cutoff length $l_m$ increases, all atoms become paired.
For a small number of atoms ($N =30$) in a given spatial configuration, we can also find the ground state numerically by matrix product state (MPS) algorithms \cite{Schollwock2011}. 
Given the MPS ground state, we calculate the projection into the singlet state $\bra{S}\hat{\rho}_{ij}\ket{S}$ of  the two-atom reduced density matrix $\hat{\rho}_{ij}$ of atoms $i,j$,  and identify pairing if the projection is the largest compared to any other combinations ($i, j'\neq j$ or $i'\neq i, j$). Once all pairings are identified for a given run (\textit{e.g.} in Fig. \ref{fig:med}\textbf{(b)}), we assume that such a state was formed according to the RG rules  and use the expression of $\tilde{J}_{j j'}$ found previously to assign an effective $l_m$ to each pair.
In Fig. \ref{fig:MPS} we also plot the MPS result for $10^5$ random distributions of 30 atoms on a lattice with the same parameters used in solving the RG flow equation. We see that the MPS and RG flow equations agree well. The discrepancy for small $l_m$ is attributable mostly to the fact that in the physical system and in the MPS simulations, there is a discreteness of atomic positions, which must however be approximated by a smooth distribution in order to solve the continuous differential equation of
Eq. (\ref{eq:flow}) (see Appendix).

\par
To appreciate the importance of interactions mediated by singlet pairs that have been integrated out, we consider the bond nesting structure, \textit{i.e.} the likelihood of finding a singlet pair  with  $n_l$ other pairs nested inside (\textit{e.g.} in Fig. \ref{fig:med}\textbf{(a)} $n_l=1$ for pair 1-4, as pair 2-3 is nested inside). 
We introduce the \textit{joint distribution}: $P(n_l, l, l_m)\text{d}l$, which gives the probability of finding a coupling length $l$ \textit{and} containing $n_l$ \textit{nested} bonds inside, when the shortest coupling length in the system is $l_m$. 
An RG flow equation can be obtained similar to the case of Eq. (\ref{eq:flow}) (see Appendix), and solved numerically.
We then obtain the fractions of nested bonds as a function of $l_m$, and plot them in Fig. \ref{fig:MPS}\textbf{(b)}, together with the result from MPS simulations. When $l_m$ is small most of the bonds are unnested ($n_{l_m}=0$),  indicating the direct pairing of consecutive atoms, but as $l_m$ increases the fractions of nested bonds ($n_{l_m}=$ 1  and 2) increase and eventually overtake the unnested bonds. 
The MPS result qualitatively agrees with that of RG. The noticeable shift can be attributed both to the discreteness of atomic positions, and to
 the relatively small size of the system (30 atoms) used in our simulations, as this is unfavorable to forming long-distance nested bonds.


To qualitatively understand
 the significant effect of bond renormalization, we also compare the RG result with simply identifying shortest distances between any two neighboring atoms in the system and pairing them up, without any distance renormalization or bond nesting. This bond distribution can be obtained from the solution to the joint flow equation $P(n_{l_m}=0, l_m, l_m)$. In Fig. \ref{fig:MPS}\textbf{(a)} we plot the number of unpaired atoms in this ``no RG" case. One observes that $\sim$ 15\% of atoms remain unpaired as $l_m \rightarrow \infty$, and that the nesting of RG is required to further eliminate long couplings (Fig. \ref{fig:MPS}\textbf{(b)}).


Thus far, we have described the ground-state properties of Eq. (\ref{eq:HI}). However, as this Hamiltonian is an effective one produced by external laser driving, the ground (or other low-energy) state cannot generally be reached by thermalization. Thus, we consider adiabatic evolution, under a time evolution process from the ground state of a non-interacting Hamiltonian which can be easily prepared in experiment.
To make our discussion specific, we consider a time evolution process described by the Hamiltonian
$\hat{H}(t) =\cos(\omega \, t)\hat{H}_0 + \sin (\omega \, t) \hat{H}_{int}^N$,
in which $\hat{H}_0$ is the non-interacting Hamiltonian whose ground state is the initial state at $t=0$, and $\hat{H}_{int}^N$ is the interacting Hamiltonian whose ground state one wants to reach at $t=\pi/2\omega$. The slew rate $\omega$ characterizes how fast the time evolution happens.
In order for this procedure to work, one needs to choose $\hat{H}_0$ in a way that $\hat{H}(t)$ avoids extra conserved (or nearly conserved) quantities, which would prevent the initial state from evolving to the final, interacting ground state. We find that a good candidate consists of an effective magnetic field, whose orientation 
 rotates in the $x$-$y$ plane by a fixed angle from site to site, and the Hamiltonian takes the form:
$\hat{H}_0 =   \epsilon_0\sum_{i} \hat{\sigma}^{i}_{\perp}(\phi_i) $,
with
$\hat{\sigma}^{i}_{\perp}(\phi_i) =\cos \phi_i \, \hat{\sigma}^{i}_{x} + \sin \phi_i \, \hat{\sigma}^{i}_{y} $
and $\phi_i=(x_i/a) \, \phi_0$, where $\phi_0$ is a constant angle between 0 and $2\pi$ chosen by the experiment. The corresponding initial state of atom $i$ is then given by
$
\left(1/\sqrt{2}\right) \left(\ket{\uparrow}-e^{i\phi_i}\ket{\downarrow}\right).
$


In practice, the optimal rate $\omega$ will be dictated by a balance of evolving slowly enough to preserve adiabaticity, and fast enough to avoid realistic errors not captured by $\hat{H}_{int}^N$, which in this case consist of losses of the photonic crystal and the spontaneous emission of photons by atoms into free space. 
We first discuss the errors associated with non-adiabaticity, which causes the final state to end up in an excited state of  $\hat{H}_{int}^N$.
Although the scaling of errors vs. slew rate $\omega$ is generally complicated for a many-body system \cite{Polkovnikov2011}, here, we can develop a simple picture based on the observation that the ground state consists of singlet pairs. The Landau-Zener theorem \cite{Landau1932, LandauLifshitz, Zener1932} then implies that a singlet will form provided that the renormalized interaction strength  between two atoms exceeds the slew rate ($\tilde{J}_{ij}\gtrsim \omega$).

\par
\begin{figure}[h]
\centering
\includegraphics[scale=0.5]{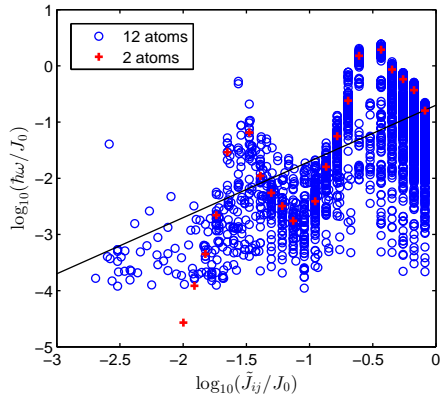}
\caption{Slew rate at pair breaking vs. binding energy for 1,000 random distributions of 12 atoms on 100 trapping sites (blue circles). The relevant parameters used in the simulations are $\epsilon_0=J_0$,
 $L=5\, a$, and $\phi_0=\pi/6$. The result for 2 atoms is also shown (red crosses). The black line is a guide to the eye,  showing a $\omega \propto \tilde{J}_{ij}$ scaling.
}
\label{fig:scal}
\end{figure}

In order to have a large proportion of long bond lengths, we choose a relatively low filling fraction and numerically simulate the time evolution of 12 atoms randomly distributed among 100 lattice sites (see caption in Fig. \ref{fig:scal} for specific parameters). In each evolution run we start from the ground state of the Hamiltonian $\hat{H}_0$ as the initial state $\psi(t=0)$, and evolve it to $\psi(t=\pi/2\omega)$ at some slew rate $\omega$. We vary the speed $\omega$ from run to run and record the pair breaking for each run (see Appendix for more information). In Fig. \ref{fig:scal}, we plot the slew rate at which a pair of atoms ($i$ and $j$) is broken vs. the effective interaction strength of the pair $\tilde{J}_{ij}$, repeated over 1000 random distributions. As a guide to the eye, we also plot the scaling $\omega \propto \tilde{J}_{ij}$, as would be expected from the simple Landau-Zener argument.
The full numerics appears consistent with this simple argument, albeit with a large spread and some oscillatory behavior.
 The oscillation is an effect that arises even in the problem of $N=2$ atoms, as the specific distance of separation gives rise to a different effective field direction and $\hat{H}_0$. To confirm this, in Fig. \ref{fig:scal} we also plot in red the result for $N=2$ atoms, which indeed exhibits the same oscillations.
The additional large variation seen for $N=12$ atoms arises from the combination of many-body effects and sampling over many random configurations.
 This inevitably results in certain configurations (\textit{e.g.} three atoms occupying consecutive sites) where renormalization group cannot quantitatively capture the full microscopic physics.
Having obtained this scaling of $\omega$-$\tilde{J}_{ij}$, next we take into account the errors in realistic experiments due to photon losses of the PCW and atomic spontaneous emission \cite{Douglas2015}. We estimate that (see Appendix) for a system of 12\% filling fraction and with the same parameters given in Fig. \ref{fig:scal},  approximately 70\% of all atoms will remain paired at the end of the time evolution when $\omega$ is optimized. 

\par
There are at least two scenarios where the RSP can be experimentally realized: one is based upon the current loading techniques described in ref. \cite{Hood2016},  where the atomic positions are truly unknown from shot-to-shot, and only global measurements are possible. Another possibility is the integration of a tweezer array \cite{Endres2016, Barredo2016, Lee2017, Barredo2018, Brown2019} with PhC's \cite{Samutpraphoot2019}, which allows disordered configurations to be created in a deterministic fashion, and individual measurements of atoms at the ends. In the former case, there is no clear way to measure the microscopic properties of entanglement (\textit{e.g.} nesting), however one can attempt to measure global spin properties instead. In particular, measurements of the collective spin operators $\hat{S}_{\alpha}=(\hbar/2)\sum_i \hat{\sigma}_{\alpha}^i$, 
where $\alpha=x, y, z$ should exhibit $<\hat{S}_{\alpha}>= 0$ and variance $< \Delta \hat{S}_{\alpha}>=0$ for a state composed globally of singlets. 
We note  $\hat{S}_{\alpha}$ can be measured by standard quantum nondemolition  (QND) techniques for atom-light interfaces \cite{Hammerer2010}, such as through the collective coupling of atoms to the guided modes of the PhC.


\textit{Conclusion} 
We have shown that the combination of long-range interactions and low filling fraction in atom-PhC interfaces can give rise to novel many-body phases typically associated to short-range, disordered systems, specifically a random singlet phase. We have also discussed how such states can be realistically prepared and measured. As such interfaces are in principle quite tunable, in the range of interaction, the dimensionality (1D or 2D) and the form of the spin interaction \cite{Chang2018}, it would be interesting in the future to explore other phenomena associated with random systems, such as many-body localization \cite{Abanin2019}, spin glasses \cite{Mydosh2015}, or the formation of large effective spins \cite{Westerberg1995}.


\textit{Acknowledgement} 
The authors are grateful to H.J. Kimble for helpful discussions. The authors acknowledge support from ERC Starting Grant FOQAL, MINECO Severo Ochoa Grant SEV-2015-0522, CERCA Programme/Generalitat de Catalunya, Fundació Privada Cellex, Fundació Mir-Puig, QuantumCAT (funded within the framework of the ERDF Operational Program of Catalonia), and Plan Nacional Grant ALIQS, funded by MCIU, AEI, and FEDER.


\appendix*

\setcounter{equation}{0}
\setcounter{figure}{0}

\renewcommand{\theequation}{A\arabic{equation}}
\renewcommand{\thefigure}{A\arabic{figure}}


\section{}


\subsection{Derivation of the renormalized Hamiltonian}
Here, starting from the system Hamiltonian of Eq. (\ref{eq:HI}) in the main text, we derive the effective Hamiltonian that results from integrating out the strongest interacting pair of atoms, which then gives $\tilde{J}_{j j'}$ in the main text. 

\par
Let's denote the indices of the spins with the strongest coupling as 1 and 2, and $j \neq 1, 2$ denote all other spins. The full Hilbert space spanned by all the spins can be divided into a low-energy subspace spanned by $\ket{S}_{12} \otimes \{ \ket{\sigma_j}, j\neq 1, 2 \}$, where $\ket{S}_{12}$ is the singlet state of spin 1 and 2, and $\sigma_j=\uparrow$ or $\downarrow$, and a high-energy subspace spanned by $\ket{T}_{12}^{0,\pm 1} \otimes \{ \ket{\sigma_j}, j\neq 1, 2 \}$, where $\ket{T}_{12}^{0,\pm 1}$ is the triplet manifold with the magnetic quantum number of each state denoted explicitly. Accordingly, the full Hamiltonian of the system can be split up into the form:
\begin{equation}\label{eq:Hsplit}
\hat{H}_{int}^N=\hat{H}_{12}+\sum_{\substack{i=1, 2 \\ j\neq 1, 2}} \hat{H}_{ij} + \sum_{j<j'\neq 1, 2}\hat{H}_{j j'} \, ,
\end{equation}
where $\hat{H}_{i j}= (J_{ij}/2) (\hat{\sigma}_{x}^i\hat{\sigma}_{x}^j + \hat{\sigma}_{y}^i\hat{\sigma}_{y}^j)$. The second term on the right-hand side of Eq. (\ref{eq:Hsplit}) describes the interactions between atom 1 and 2 with all the other atoms, while the third term describes the interactions between all the atoms excluding 1 and 2, and henceforth we will call them $\hat{V}_{\text{od}}$ and $\hat{V}_{\text{d}}$ respectively. Using the ``Schrieffer-Wolff" transformation \cite{Bravyi2011}, we can obtain an effective Hamiltonian governing the dynamics of the remaining spins $j\neq 1,2$ in the low-energy subspace, given by
\begin{equation}
\hat{H}_{int}^N=-J_{12} + \frac{1}{2}\hat{P}_0 \left[ \hat{S}, \hat{V}_{\text{od}}\right] \hat{P}_0 + \hat{V}_{\text{d}} \, ,
\end{equation}
where $\hat{P}_0=\ket{S}_{12}\bra{S}_{12} \otimes \mathbb{I}$  projects a state into the low-energy subspace, and $\hat{S}=\sum_{p, q}\frac{\bra{p}\hat{V}_{\text{od}}\ket{q}}{E_p-E_q}\ket{p}\bra{q}$ with $p$ and $q$ denoting states belonging to different subspaces. After some algebra, we reach
\begin{equation}
\hat{H}_{int}^N=-J_{12} - \sum_{j\neq 1, 2}\frac{\left( J_{2j}-J_{1j}\right)^2}{2J_{12}} + \hat{V}_{\text{d}}' \, ,
\end{equation}
with $\hat{V}_{\text{d}}'$ takes the same form as $\hat{V}_{\text{d}}$, but where the bare coupling strength $J_{j j'}$ is replaced by the renormalized value $\tilde{J}_{j j'}=J_{j j'}-\frac{\left( J_{2j}-J_{1j}\right)\left( J_{2j'}-J_{1j'}\right)}{J_{12}}$. Plugging in the expression $J_{ij}=J_0 \, \exp(-|x_i-x_j|/L)$, one can show that when atom $j$ and $j'$ sit on the \textit{same} side of atoms 1 and 2, $\tilde{J}_{j j'}\approx J_{j j'}$, whereas when they sit on \textit{opposite} sides of atoms 1 and 2, $\tilde{J}_{j j'}=J_0 \, \exp\left(-\frac{|x_j-x_{j'}|-d_{\text{eff}}}{L}\right)$, with the expression of $d_{\text{eff}}$ given in the main text. Thus, after the two spins with the strongest coupling in the system form into a singet pair, the form of interactions between remaining spins sitting on opposite sides of spin 1 and 2 stays the same, except their distances are renormalized and shrunk by an amount of $d_{\text{eff}}$.


\subsection{Renormalization group flow equations}
\subsubsection{Derivation of the flow equation for the coupling length distribution}
Here, we describe further the derivation of the renormalization group Eq. (\ref{eq:flow}) of the main text, and discuss its solution. During the RG process, after all pairs of interacting atoms separated by an effective distance of $l_m$ have been integrated out, the remaining distances $l$ follow a probability distribution $P(l,l_m) \text{d}l$. We next consider how $P(l,l_m)$ evolves when $l_m$ is increased by an infinitesimal amount $\Delta$:
\begin{multline}
P(l,l_m+\Delta)=\left[1-2\Delta P(l_m, l_m)\right]^{-1}\big[ P(l, l_m)-2\Delta P(l_m, l_m)P(l, l_m) \int_{l_m}^\infty \text{d}xP(x, l_m) \\
+\Delta P(l_m, l_m) \int_{l_m}^\infty \text{d}x \, \text{d}y \, P(x, l_m) P(y, l_m) \, \delta(x+y+l_m-d_{\text{eff}}-l) \big] \, , \label{eq:flowA}
\end{multline}
The term in the first square bracket on the RHS of the equation takes care of the normalization of the probability (sum is 1). The second term in the second  square bracket on the right-hand side corresponds to all the couplings with effective distance $l$ that disappear when all the atomic pairs with effective distance $l_m$ are eliminated (see Fig. \ref{fig:flow}\textbf{(a)} below), while the  third term corresponds to all the pairs with effective distance $l$ that newly appear when all the pairs with effective distance $l_m$ are eliminated (see Fig. \ref{fig:flow}\textbf{(b)} below).
\begin{figure}[h]
\includegraphics{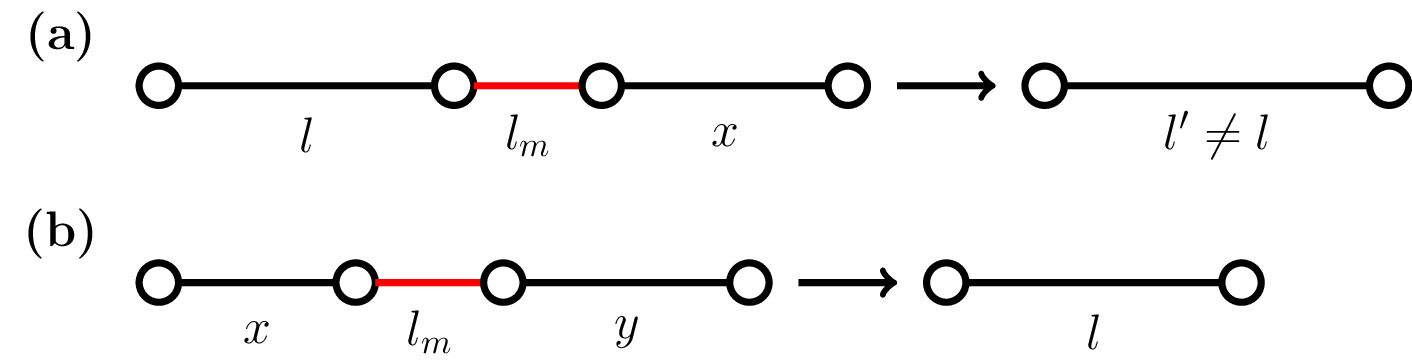}
\caption{Illustration of various terms in Eq. (\ref{eq:flow}). The atoms involved in the RG are denoted by circles. The strongest interacting pair, separated by effective distance $l_m$ (red), is integrated out, leading to new effective distances between the remaining pair of atoms.}
\label{fig:flow}
\end{figure}
By performing the change of variables $\lambda=l/l_m-1$ and $Q(\lambda, l_m)=l_m P(l,l_m)$, after some algebra, one reaches Eq. (\ref{eq:flow}) in the main text.


\subsubsection{Initial condition for the flow equation}
We assume that atoms are randomly distributed among the sites of the PCW, with probability $P = N/N_{\text{site}}$, under the conditions $N, N_{\text{site}} \rightarrow \infty$, where $N$ and $N_{\text{site}}$ are the total number of atoms and lattice sites, respectively. Then, starting from any occupied site in the spin chain, the probaiblity of finding the next atom at the $n$ th next site away is
\begin{equation}\label{eq:Pi}
\mathcal{P}\left( l=na, l_m=a\right)=P(1-P)^{n-1}.
\end{equation} 
This is equivalent to the probability of finding two atoms positioned $l=na$ apart, when the smallest distance between two atoms in the whole system is $l_m =a$, with $a$ being the lattice constant. We note the normalization condition that Eq. (\ref{eq:Pi}) fulfills:
\begin{equation}
\sum_{n=1}^{+\infty} P(1-P)^{n-1} = 1.
\end{equation}

To line up the solution of the flow equation at the beginning as much as possible with this discrete distribution, we may approximate it by a continuous distribution:
\begin{equation}
\mathcal{P}\rightarrow P(1-P)^{{l/l_m-1}} \frac{\text{d}l}{l_m} \equiv P(l,l_m) \, \text{d}l  \, ,
\end{equation}
where $l$ is a continuous variable and $P(l,l_m)$ is a \textit{probability density}. Enforcing normalization condition  
$
\int_{l_m}^{+\infty} P(l,l_m) \, \text{d}l = 1
$
we obtain the initial condition for the flow equation Eq. (\ref{eq:flow}) in the main text:
\begin{equation} \label{eq:ini}
Q(\lambda, l_m= a)=-\ln (1-P) (1-P)^{\lambda} \, .
\end{equation}

\subsubsection{Derivation of the joint flow equation}

Similar as for $P(l, l_m)$, a flow equation for the nested probability distribution
$P(n_l, l, l_m)$ can be constructed as
\begin{multline}
P(n_l, l, l_m+\Delta)\left(1-2\Delta\sum_{n_{l_m}=0}^{+\infty} P(n_l, l_m, l_m)\right)=
P(n_l, l, l_m)-2\Delta P(n_l, l, l_m) \sum_{n_{l_m}=0}^{+\infty} P(n_l, l_m, l_m) \\
+ \Delta\cdot\sum_{n_{l_m},n_x,n_y=0}^{+\infty}  \delta_{n_x+n_{l_m}+n_y+1,n_l}  P(n_{l_m}, l_m, l_m) \, \times  \\
\int_{l_m}^{+\infty}\text{d}x \, \text{d}y \, P(n_x, x, l_m) P(n_y, y, l_m) \delta \left[x+y+l_m-d_{\text{eff}}(l_m)-l\right] \, ,
\end{multline}
where the terms have similar meanings as those in Eq. (\ref{eq:flowA}). We note $n_l$ can take any non-negative interger values $n_l=0, 1, 2, ...$ and initially when $l_m=a$, $n_l=0$ for all $l$.

Again, introducing the substitutions $\lambda =l/l_m-1$ and $P(n_l, l, l_m) \text{d}l =Q(n_{\lambda}, \lambda, l_m) \text{d}\lambda$, after some algebra, one obtains
\begin{multline}\label{eq:jflow}
l_m\frac{\partial Q}{\partial l_m} - (1+\lambda)\frac{\partial Q}{\partial \lambda} =Q 
+\sum_{n_0=0}^{n_{\lambda}-1}Q(n_0,0,l_m) \sum_{n_x=0}^{n_{\lambda}-1-n_0} \\
\times \int_0^{\lambda+g(l_m)}  \text{d}\lambda_x Q(n_x,\lambda_x , l_m) Q(n_{\lambda}-1-n_0-n_x,\lambda+g(l_m)-\lambda_x, l_m) \, .
\end{multline}

\subsection{Simulations of the time evolution}

In this section, we describe in detail the simulations we did to determine the relation between the slew rate at which a singlet pair of atoms breaks and the effective bond strength of the pair.  We start from the ground state of the non-interacting Hamiltonian $\hat{H}_0$ as the initial state $\psi(t=0)$, evolve it under the time-dependent Hamiltonian $\hat{H}(t)$ at some chosen slew rate $\omega$, and at the end of the evolution we compare the final state $\psi(t=\pi/2\omega)$ with the true ground state of $\hat{H}_{int}^N$, obtained by direct diagonalization. For each distribution of 12 atoms among 100 sites, we begin with a small $\omega$ at which our final state will have a high overlap with the true ground state of $\hat{H}_{int}^N$ (\textit{e.g.} 99\%). Here, all atoms form singlet pairs, which are numerically identified by looking at the singlet fraction in two-atom reduced density matrices. We then  repeat the time evolution at faster slew rates, recording the value $\omega$ at which each singlet bond breaks (defined as the singlet fraction in the final state dropping below 50\%). Separately, from the bond nesting pattern of the true ground state, we can identify the effective, renormalized coupling strength $\tilde{J}_{ij}$ associated with each pair. We then plot the slew rate at bond breaking vs. the coupling strength for each pair, as in Fig. 4 in the main text.

\subsection{Errors due to photon losses in experiments and optimization of the slew rates}

In this section we take into accont the errors in realistic experiments due to photon losses of the PCW and atomic spontaneous emission, and find the optimal slew rate that will preserve the most singlet pairs at the end of the time evolution. 
In ref. \cite{Douglas2015}, it was shown that when the system is optimized, the photon loss processes give rise to incoherent spin flips at a rate given by $\sim J_0/\sqrt{C}$, where $C$ is the single atom cooperativity in the PCW. 
Since the total evolution time is $T=\pi/2\omega$, the probability that a singlet is lost incoherently is then $P_{\text{inc}}(\omega)=J_0\,T/\sqrt{C}$. 
Without accounting for this incoherent loss, the errors would be purely due to non-adiabacity, and the fraction of \textit{unpaired} atoms at the end of the time evolution due to finite slew rate: $F_{\text{unpaired}}(\omega)$ can be found by using the scaling between $\omega$ and $\tilde{J}_{ij}$ and the relation between the effective coupling length $l_m$ and the fraction of unpaired atoms (\textit{e.g.} as plotted in Fig. 3\textbf{(a)}). Then taking into account incoherent losses, the fraction of \textit{paired} atoms at the end of the time evolution can be obtained as $F_{\text{paired}}(\omega)=(1-F_{\text{unpaired}}(\omega))(1-P_{\text{inc}}(\omega))$. Taking a PCW cooperativity of $C=10^{4}$  \cite{Douglas2015} in $P_{\text{inc}}$, and optimizing over $\omega$ for maximum $F_{\text{paired}}(\omega)$, we estimate that approximately $F_{\text{paired}}(\omega)\approx$ 70\% for 12\% filling fraction.


\bibliography{PhC}


\end{document}